\documentclass[smus]{snow2e}
\usepackage{epsfig}
\begin{document}
\title{Inclusive Production of $t\bar{t}$ Pairs in Hadronic Collisions\thanks
{Contribution to the 1996 DPF/DPB Summer Study on New Directions for High 
Energy Physics, Snowmass, June - July, 1996.  Argonne report ANL-HEP-CP-96-95. 
This work was supported by the U.S. Department of Energy, Division of High 
Energy Physics, Contract No. W-31-109-ENG-38.}}

\author{Edmond L. Berger and Harry Contopanagos\\ {\it High Energy Physics 
Division, Argonne National Laboratory, Argonne, IL 60439, USA}}

\maketitle

\thispagestyle{empty}\pagestyle{empty}

\begin{abstract}
We summarize our calculation of the total cross section for top quark
production at hadron colliders within the context of perturbative quantum
chromodynamics, including resummation of the effects of initial-state
soft gluon radiation to all orders in the strong coupling strength.
\end{abstract}

\section{Introduction and Motivation}

In hadron interactions at collider energies, $t\bar{t}$ pair production 
proceeds through partonic hard-scattering processes involving initial-state
light quarks $q$ and gluons $g$.  In lowest-order perturbative
quantum chromodynamics (QCD), ${\cal O}(\alpha_s^2)$,  the two partonic 
subprocesses 
are $q + \bar{q} \rightarrow t + \bar{t}$ and $g + g \rightarrow t + \bar{t}$.  
Calculations of the cross section through next-to-leading order, 
${\cal O}(\alpha_s^3)$, involve gluonic radiative corrections to these 
lowest-order subprocesses as well as contributions from the $q + g$ initial 
state~\cite{ref:dawson}.  A complete fixed-order calculation at order 
${\cal O}(\alpha_s^n), n \ge 4$ does not exist.  

The physical cross section for each production channel is obtained through 
the convolution
\begin{equation}
\sigma_{ij}(S,m)=
{4m^2\over S}\int_0^{{S\over 4m^2}-1}d\eta
\Phi_{ij}(\eta,\mu) \hat\sigma_{ij}(\eta,m,\mu) .
\label{feleven}
\end{equation}
%\begin{eqnarray}
%&\sigma_{ij}(S,m)=
%{4m^2\over S}\int_0^{{S\over 4m^2}-1}d\eta&              \nonumber \\
%& \times \Phi_{ij}\biggl[{4m^2\over S}(1+\eta),\mu^2\biggr]
%\hat\sigma_{ij}(\eta,m^2,\mu^2)& .
%\label{feleven}
%\end{eqnarray}
%
The square of the total hadronic center-of-mass energy is $S$, the square of 
the partonic center-of-mass energy is $s$, $m$ denotes the top mass, $\mu$ is 
the usual factorization and renormalization scale, and $\Phi_{ij}(\eta,\mu)$ is 
the parton flux.  
The variable $\eta={s \over 4m^2} - 1$ measures the distance from the 
partonic threshold.  The indices $ij\in\{q\bar{q},gg\}$ denote the initial 
parton channel.  The partonic cross section 
$\hat\sigma_{ij}(\eta,m,\mu)$ is obtained commonly from fixed-order QCD
calculations~\cite{ref:dawson}, or, as described here, from calculations
that go beyond fixed-order perturbation theory through the inclusion of 
gluon resummation~\cite{ref:laeneno,ref:edpapero,ref:catani} to all orders in 
the strong coupling strength $\alpha_s$.  We use the notation 
$\alpha \equiv \alpha(\mu=m) \equiv \alpha_s(m)/\pi$.  The total physical 
cross section is obtained after incoherent addition of the contributions from 
the the $q\bar{q}$ and $gg$ production channels.  

Comparison of the partonic cross section at next-to-leading order with its 
lowest-order value reveals that the ratio becomes very large in the 
near-threshold region.  Indeed, as $\eta \rightarrow 0$, the ``$K$-factor" at 
the partonic level $\hat K(\eta)$ grows in proportion to $\alpha \ln^2(\eta)$. 
The very 
large mass of the top quark notwithstanding, the large ratio $\hat K(\eta)$ 
makes it evident that the next-to-leading order result does not necessarily 
provide a reliable quantitative prediction of the top quark production cross 
section at the energy of the Tevatron collider.  The large ratio casts doubt 
on the reliability of simple fixed-order perturbation theory for physical 
processes for which the near-threshold region in the subenergy variable 
contributes significantly to the physical cross section.  Top quark production 
at the Fermilab Tevatron is one such process, because the top mass is 
relatively large compared to the energy available.  Other examples include 
the production of hadronic jets that carry large values of transverse momentum 
and the production of pairs of supersymmetric particles with large mass.  To 
obtain more reliable theoretical estimates of the cross section in 
perturbative QCD, it is important first to identify and isolate the terms that 
provide the large next-to-leading order enhancement and then to resum these 
effects to all orders in the strong coupling strength.  

\section{Gluon Radiation and Resummation}

The origin of the large threshold enhancement may be traced to initial-state
gluonic radiative corrections to the lowest-order channels.  
We remark that we are calculating the inclusive total cross 
section for the production of a top quark-antiquark pair, i.e., the total 
cross section for $t + \bar{t} + \rm anything$.  The partonic subenergy 
threshold in question is the threshold for $t + \bar{t} +$ any number of 
gluons.  This coincides with the threshold in the invariant mass of the 
$t + \bar{t}$ system for the lowest order subprocesses only. 

For $i + j \rightarrow t + \bar{t} + g$, we define the variable $z$ through 
the invariant
$(1-z) = {2k \cdot p_t \over m^2}$, where $k$ and $p_t$ are the four-vector 
momenta of the gluon and top quark.  In the limit that 
$z \rightarrow 1$, the radiated gluon carries zero momentum.  After cancellation
of soft singularities and factorization of collinear singularities in 
${\cal O}(\alpha_s^3)$, there is a left-over integrable large logarithmic 
contribution to the partonic cross section associated with initial-state gluon 
radiation.  This contribution is often expressed in terms of ``plus" 
distributions.  In ${\cal O}(\alpha_s^3)$, it is proportional to 
$\alpha^3 \ln^2(1-z)$.  When integrated over the near-threshold 
region $1 \ge z \ge 0$, it provides an excellent approximation to the full 
next-to-leading order physical cross section as a function of the top mass.  
At $m =$ 175 GeV, the ratio of the next-to-leading order to the leading order 
physical cross sections in the leading logarithmic approximation is 
$\sigma_{q\bar{q}}^{(0+1)}/\sigma_{q\bar{q}}^{(0)} = $1.22.  This ratio shows 
that the near-threshold logarithm builds up cross section in a worrisome 
fashion.  It suggests that perturbation theory is not converging to a stable 
prediction of the cross section.   The goal of gluon resummation is to sum the 
series in $\alpha^{n+2} \ln^{2n}(1-z)$ to all orders in $\alpha$ in order to 
obtain a more defensible prediction.

Different methods of resummation differ in theoretically and phenomenologically
important respects.  Formally, if not explicitly in some approaches, an 
integral over the radiated gluon momentum $z$ must be done over regions in 
which $z \rightarrow 1$.  Therefore, one significant distinction among methods 
has to do with how the inevitable ``non-perturbative" region is handled.  In 
the approach of Laenen, Smith, and van Neerven (LSvN)~\cite{ref:laeneno}, an 
undetermined infrared  cutoff (IRC) $\mu_o$ is introduced, 
with $\Lambda_{QCD} \leq \mu_o \leq m$.  
The presence of an extra scale spoils the renormalization 
group properties of the overall expression.  The unfortunate dependence of the 
resummed cross section on this undetermined cutoff is important numerically 
since it appears in an exponent~\cite{ref:laeneno}.  It is difficult to evaluate
theoretical uncertainties in a method that requires an undetermined
infrared cutoff.  
\begin{figure}[b]
\leavevmode
\begin{center}
\epsfig{figure=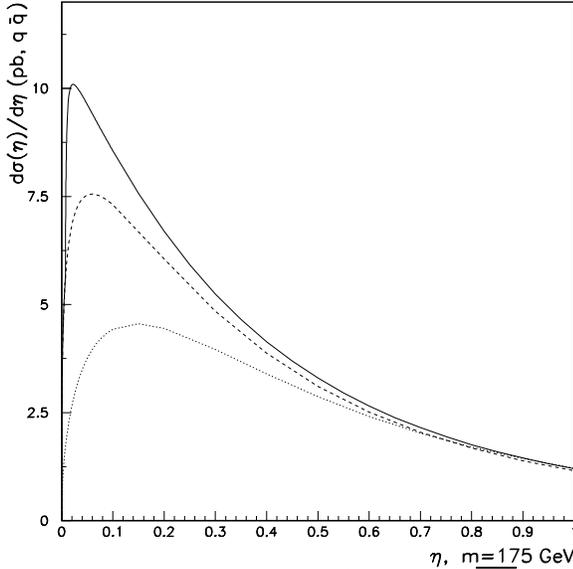,height=2.75in}
\end{center}
\vskip -0.9cm
\caption{Differential cross section $d\sigma/d\eta$ in the
${\overline{\rm MS}}$ scheme for the $q\bar{q}$  channel:
Born (dotted), next-to-leading order (dashed) and resummed (solid).}
\label{fig:fone}
\end{figure}
\section{Perturbative Resummation}

The method of resummation we employ~\cite{ref:edpapero} is based on a 
perturbative truncation of principal-value resummation~\cite{ref:stermano}.  
This approach has an important technical advantage in that it 
does not depend on arbitrary infrared cutoffs.  Because extra scales are 
absent, the method permits an evaluation of its perturbative regime of
applicability, i.e., the region of the gluon radiation phase
space where perturbation theory should be valid.  We work in the 
$\overline{\mbox{MS}}$ factorization scheme.

Factorization and evolution lead directly to exponentiation of the set of 
large threshold logarithms in moment ($n$) space in terms of an exponent 
$E^{PV}$:
\begin{equation}
E^{PV}(n, m^2)\equiv -\int\limits_P
d\zeta {{\zeta^{n-1}-1}\over{1-\zeta}}
\int\limits^1_{(1-\zeta)^2}
{{d\lambda}\over{\lambda}}g\left[\alpha\left(\lambda m^2\right)\right].
\label{bseven}
\end{equation}
The function $g(\alpha)$ is calculable perturbatively.  All large soft-gluon 
threshold contributions are included through the two-loop running of $\alpha$.
The integral in the complex plane runs along a contour $P$ with endpoints 0 
and 1 that is symmetric under reflections across the real axis.  

The function $E^{PV}$ is finite, and 
$\lim_{n\rightarrow\infty}E^{PV}(n,m^2)=-\infty$.  Therefore, 
the corresponding partonic cross section is finite as $z\rightarrow 1
\ (n\rightarrow +\infty)$.
The function $E^{PV}$ includes both perturbative and non-perturbative content.  
The non-perturbative content is not a prediction of perturbative QCD. 
We choose to use the exponent only in the interval in moment space in which the 
perturbative content dominates.
We use the attractive finiteness of Eq.~(\ref{bseven}) to derive a 
perturbative asymptotic representation of $E(x,\alpha(m))$ that is 
valid in the moment-space interval
\begin{equation}
1<x\equiv \ln n< t\equiv {1\over 2\alpha b_2}.
\label{tseven}
\end{equation}
The coefficient $b_2=(11C_A-2n_f)/12$; the number of flavors $n_f=5$; 
$C_{q\bar{q}}=C_F=4/3$; and $C_{gg}=C_A=3$. 

The perturbative asymptotic representation is
\begin{equation}
E_{ij}(x,\alpha)\simeq E_{ij}(x,\alpha,N(t))=
2C_{ij}\sum_{\rho=1}^{N(t)+1}\alpha^\rho
\sum_{j=0}^{\rho+1}s_{j,\rho}x^j\ .
\label{teight}
\end{equation}
Here
\begin{equation}
s_{j,\rho}=-b_2^{\rho-1}(-1)^{\rho+j}2^\rho c_{\rho+1-j}(\rho-1)!/j!\ ;
\label{tnine}
\end{equation}
and $\Gamma(1+z)=\sum_{k=0}^\infty c_k z^k$, where $\Gamma$ is the Euler gamma 
function.  
The number of perturbative terms $N(t)$ in Eq.~(\ref{teight}) is
obtained~\cite{ref:edpapero} by optimizing the asymptotic approximation
$\bigg|E(x,\alpha)-E(x,\alpha,N(t))\bigg|={\rm minimum}$. 
Optimization works perfectly, with $N(t)=6$ at $m = 175$ GeV.   
As long as $n$ is in the interval of Eq.~(\ref{tseven}),
all the members of the family in $n$ are optimized 
at the same $N(t)$, showing that the optimum number of 
perturbative terms is a function of $t$, i.e., of $m$ only.

Because of the range of validity in Eq.~(\ref{tseven}), terms in the exponent 
of the form $\alpha^k\ln^kn$ are of order unity, and terms with fewer powers
of logarithms, $\alpha^k\ln^{k-m}n$, are negligible.
Resummation is completed in a finite number of steps.  Upon using the running 
of the coupling strength $\alpha$ up to two loops only, 
all monomials of the form $\alpha^k\ln^{k+1}n,\ \alpha^k\ln^kn$
are produced in the exponent of Eq.~(\ref{teight}).  We discard monomials
$\alpha^k\ln^kn$ in the exponent because of the restricted leading-logarithm
universality between $t\bar{t}$ production and massive lepton-pair 
production, the Drell-Yan process.

The exponent we use is the truncation
\begin{equation}
E_{ij}(x,\alpha,N)=2C_{ij}\sum_{\rho=1}^{N(t)+1}\alpha^\rho s_\rho x^{\rho+1} ,
\label{tseventeen}
\end{equation}
with the coefficients
$s_\rho\equiv s_{\rho+1,\rho}=b_2^{\rho-1}2^\rho/\rho(\rho+1)$.  The number of
perturbative terms $N(t)$ is a function of only the top quark mass $m$.  This
expression contains no factorially-growing (renormalon) terms. 
It is valuable to stress that we can derive the perturbative expressions,
Eqs.~(\ref{tseven}), (\ref{teight}), and (\ref{tnine}),  without the 
principal-value prescription, although with less certitude~\cite{ref:edpapero}.

After inversion of the Mellin transform from moment space to 
the physically relevant momentum space, the resummed partonic cross sections, 
including all large threshold corrections, can be written
\begin{equation}
\hat{\sigma}_{ij}^{R;pert}(\eta,m)=
\int_{z_{min}}^{z_{max}}dz
{\rm e}^{E_{ij}(\ln({1\over 1-z}),\alpha)}
\hat{\sigma}_{ij}'(\eta,m,z) .
\label{bthreep}
\end{equation}
The leading large threshold corrections are contained in the exponent 
$E_{ij}(x,\alpha)$, a calculable polynomial in $x$.  The derivative
$\hat{\sigma}_{ij}'(\eta,m,z)=d(\hat{\sigma}_{ij}^{(0)}(\eta,m,z))/dz$,
and $\hat{\sigma}_{ij}^{(0)}$ is the lowest-order ${\cal O}(\alpha_s^2)$
partonic cross section expressed in terms of inelastic kinematic variables.
The upper limit of integration, $z_{max} < 1$, is set by the boundary
between the perturbative and non-perturbative regimes, well specified within
the context of the calculation, and $z_{min}$ is fixed by kinematics.  

Perturbative resummation probes the threshold down to
$\eta\ge \eta_0 =(1-z_{max})/2 $.  Below this value, perturbation theory, 
resummed or otherwise, is not to be trusted.  For $m$ = 175 GeV, we determine 
that the perturbative regime is
restricted to values of the subenergy greater than 1.22 GeV above the 
threshold ($2m$) in the $q{\bar q}$ channel and 8.64 GeV 
above threshold in the $gg$ channel. The difference reflects the larger
color factor in the $gg$ case.  The value 1.22 GeV is comparable to the decay 
width of the top quark.  

\section{Physical cross section}

Other than the top mass, the only undetermined scales are the QCD 
factorization and renormalization scales.  We adopt a common value $\mu$ for 
both, and we vary this scale over the interval $\mu/m\in\{0.5,2\}$ in order 
to evaluate the theoretical uncertainty of the numerical predictions.  We use 
the CTEQ3M parton densities~\cite{ref:cteq}.
A quantity of phenomenological interest is the differential cross section 
${d\sigma_{ij}(S,m^2,\eta)\over d\eta}$.  Its integral over $\eta$ is the 
total cross section.  In Fig.~1 we plot this distribution for the $q \bar{q}$
channel at $m=175$ GeV, ${\sqrt S}=1.8$ TeV, and $\mu=m$.   The full range of 
$\eta$ extends to 25, but we display the behavior only in the near-threshold 
region where resummation is important.  We observe that, at the energy of the 
Tevatron, resummation is quite significant for the $q\bar{q}$ channel.  A 
similar figure for the $gg$ channel may be found in our 
publications~\cite{ref:edpapero}.  
\begin{figure}[b]
\leavevmode
\begin{center}
\epsfig{figure=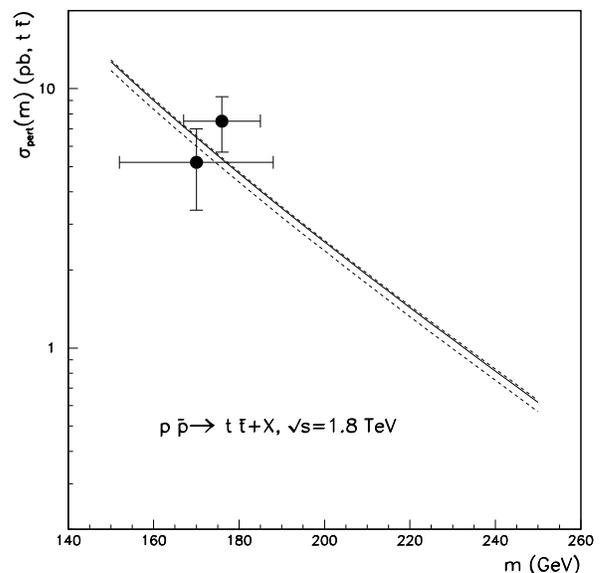,height=2.75in}
\end{center}
\vskip -0.9cm
\caption{Inclusive total cross section for top quark production in $p \bar{p}$ 
collisions at $\sqrt S$ = 1.8 TeV. The dashed 
curves show the upper and lower limits while the solid curve 
is our central prediction.  CDF and D0 data are shown.}
\label{fig:ftwo}
\end{figure}

In Fig.~2, we show our total cross section for $t\bar{t}$-production 
as a function of top mass in $p \bar{p}$ collisions 
at $\sqrt{S}=1.8$ TeV.  The central value is obtained with the 
choice $\mu/m=1$, and the lower and upper limits are  the maximum 
and minimum of the cross section in the range $\mu/m\in\{0.5,2\}$.  
At $m =$ 175 GeV, the full width of the uncertainty band 
is about 10\%\ .  We consider that the variation of the cross section over 
the range $\mu/m\in\{0.5,2\}$ provides a good overall estimate
of uncertainty.  For comparison, we note that over the 
same range of $\mu$, the strong coupling strength $\alpha$ varies by 
$\pm10$\%\ at $m$ = 175 GeV.  In estimating uncertainties, we do not consider 
explicit variations of the non-perturbative cutoff, expressed through 
$z_{max}$.  
This is justified because, for a fixed $m$ and $\mu$, $z_{max}$
is obtained by enforcing dominance of the universal leading logarithmic terms 
over the subleading ones. Therefore, $z_{max}$
is {\it derived} and is not a source of uncertainty.  At fixed $m$, the 
cutoff necessarily varies as $\mu$ and thus $\alpha$ vary. 
Using a different choice of parton 
densities~\cite{ref:mrsa}, we find a 4\%\ difference in the central value of our 
prediction~\cite{ref:edpapero} at $m =$ 175 GeV.  A comparison of the 
predictions~\cite{ref:edpapero} in the 
$\overline{\mbox{MS}}$ and DIS factorization schemes also shows a modest
difference at the level of $4\%$.

Our calculation is in agreement with the  data~\cite{ref:cdfdz}.  We find
$\sigma^{t\bar{t}}(m=175\ {\rm GeV},\sqrt{S}=1.8\ {\rm TeV})=
5.52^{+0.07}_{-0.42}\ \rm{pb}$.  This cross section is larger than the 
next-to-leading order value by about $9\%$.

The top quark cross section increases quickly with the energy of the
$p \bar{p}$ collider.  We provide predictions in Fig.~3 for an upgraded 
Tevatron operating at $\sqrt{S}=2$ TeV.  We determine 
$\sigma^{t\bar{t}}(m=175\ {\rm GeV},\sqrt{S}=2\ {\rm TeV})=
7.56^{+0.10}_{-0.55}\ \rm {pb}$.  The central value rises to 22.4 pb at 
$\sqrt{S}=3\ {\rm TeV}$ and 46 pb at $\sqrt{S}=4\ {\rm TeV}$.

Extending our calculation to much larger values of 
$m$ at ${\sqrt S}=1.8$ TeV, we find that resummation in the principal 
$q\bar{q}$ channel produces 
enhancements over the next-to-leading order cross section of $21\%$, $26\%$, 
and $34\%$, respectively, for $m =$ 500, 600, and 700 GeV.  The reason for the
increase of the enhancements with mass at fixed energy is that the threshold 
region becomes increasingly dominant.  Since the $q\bar{q}$ 
channel also dominates in the production of hadronic jets at very large values 
of transverse momenta, we suggest that on the order of $25\%$ of the excess
cross section reported by the CDF collaboration~\cite{ref:cdfjets} may well be 
accounted for by resummation.

Turning to $pp$ scattering at the energies of the Large Hadron Collider (LHC)
at CERN, we note a few significant differences from $p\bar{p}$ scattering at 
the energy of the Tevatron.  The dominance of the $q {\bar q}$ production
channel is replaced by $g g$ dominance at the LHC.  Owing to the much larger 
value of $\sqrt{S}$, the near-threshold region in the subenergy variable is 
relatively less important, reducing the significance of initial-state soft
gluon radiation.  Lastly, physics in the region of large subenergy, where 
straightforward next-to-leading order QCD is also inadequate, becomes
significant for $t\bar{t}$ production at LHC energies.  Using the approach
described in this paper, we estimate
$\sigma^{t\bar{t}}(m=175\ {\rm GeV},\sqrt{S}=14\ {\rm TeV})= $ 760 pb.

\section{Other Methods of Resummation}
Two other groups have published calculations of the total cross section at 
$m=175\ {\rm GeV}$ and $\sqrt{s}=1.8\ {\rm TeV}$:
$\sigma^{t\bar t}$({\rm LSvN}~\cite{ref:laeneno}) = $4.95^{+0.70}_{-0.40}$ pb; 
and  
$\sigma^{t\bar t}$({\rm CMNT}~\cite{ref:catani}) = $4.75^{+0.63}_{-0.68}$ pb.  
From the purely numerical point of view, all agree 
within their estimates of theoretical uncertainty.  However, the resummation 
methods differ as do the methods for estimating uncertainties.  Both 
the central value and the band of uncertainty of the LSvN predictions are 
sensitive to their arbitrary infrared cutoffs.  To estimate theoretical 
uncertainty, we use the standard $\mu$-variation, whereas LSvN obtain theirs 
primarily from variations of their cutoffs. 

The group of Catani, Mangano, Nason, and Trentadue (CMNT)~\cite{ref:catani} 
calculate a central value of 
the resummed cross section (also with $\mu/m = 1$) that is less than 
$1\%$ above the exact next-to-leading order value.  
There are similarities and differences between our approach 
and the method of CMNT.  We use the same universal leading-logarithm 
expression in moment space, but differences occur after the transformation to
momentum space.  The differences can 
be stated more explicitly if we examine the perturbative expansion of the
resummed hard kernel ${\cal H}^{R}_{ij}(z,\alpha)$.   
If, instead of 
restricting the resummation to the universal leading logarithms only, we were 
to use the full content of ${\cal H}^{R}_{ij}(z,\alpha)$, we would arrive at 
an analytic expression that is equivalent to the numerical inversion of CMNT, 
\begin{equation}
{\cal H}^{R}_{ij} \simeq 1+2\alpha C_{ij} 
\biggl[\ln^2 (1-z) + 2\gamma_E \ln (1-z)\biggr] + {\cal O}(\alpha^2).
\label{padovao}
\end{equation} 
In terms of this expansion, in our work we retain only the leading term 
$\ln^2 (1-z)$ at order 
$\alpha$, but CMNT retain both this term and the subleading term 
$ 2\gamma_E \ln (1-z)$.  Indeed, if the subleading 
term $ 2\gamma_E \ln (1-z)$ is discarded in Eq.~(\ref{padovao}), the
residuals $\delta_{ij}/\sigma_{ij}^{NLO}$ defined by CMNT~\cite{ref:catani} 
increase from 
$0.18\%$ to $1.3\%$ in the $q\bar{q}$ production channel and from $5.4\%$ to 
$20.2\%$ in the $gg$ channel.  After addition of the two 
channels, the total residual $\delta/\sigma^{NLO}$ grows from the negligible 
value 
of about $0.8\%$ cited by CMNT to the value $3.5\%$.  While still smaller than 
the increase of about $9\%$ that we obtain, the increase of $3.5\%$ vs. $0.8\%$ 
shows the substantial influence of the subleading logarithmic terms retained
by CMNT. 

We judge that it is not appropriate to keep the
subleading term for several reasons: it is not universal; it is not the same 
as the subleading term in the exact ${\cal O}(\alpha^3)$ calculation; and it
can be changed arbitrarily if one elects to keep non-leading terms in moment
space.  The subleading term is negative, and it is numerically very 
significant when integrated throughout the phase space.  
In the $q\bar{q}$ channel at $m=175$ GeV and ${\sqrt S}=1.8$ TeV, 
its inclusion eliminates more than half of the contribution from the leading 
term.  In our view, the presence of numerically significant subleading 
contributions begs the question of consistency.  The
influence of subleading terms is amplified at higher 
orders where additional subleading structures occur in the CMNT 
approach with significant numerical coefficients proportional to $\pi^2$, 
$\zeta(3)$, and so forth. We will present a more detailed discussion of these 
points elsewhere.

Our theoretical analysis and the stability of our cross section under 
variation of the hard scale $\mu$ provide confidence that our perturbative 
resummation 
yields an accurate calculation of the inclusive top quark cross section at
Tevatron energies and exhausts present understanding of the perturbative
content of the theory.
\begin{figure}[b]
\leavevmode
\begin{center}
\epsfig{figure=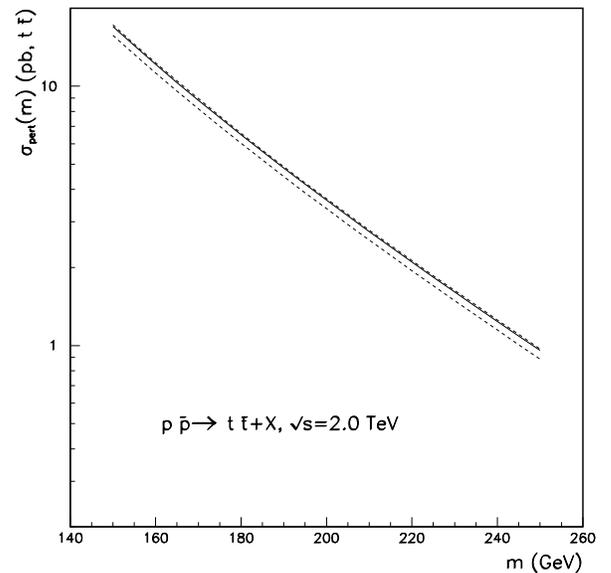,height=2.75in}
\end{center}
\vskip -0.9cm
\caption{Inclusive total cross section for top quark production 
in $p \bar{p}$ collisions at $\sqrt S$ = 2.0 TeV.}
\label{fig:fthree}
\end{figure}


\begin{thebibliography}{2}

\bibitem{ref:dawson}
P. Nason, S. Dawson, and R.~K.~Ellis, Nucl. Phys. {\bf B303}, 607 (1988);
{\bf B327}, 49 (1989); {\bf B335}, 260(E) (1990).
W. Beenakker {\it et al.}, Phys. Rev. {\bf D40}, 54 (1989); 
Nucl. Phys. {\bf B351}, 507 (1991).
\bibitem{ref:laeneno}
E. Laenen, J. Smith, and W.L. van Neerven, Nucl. Phys. {\bf B369}, 543 (1992);
Phys. Lett. B {\bf 321}, 254 (1994).
\bibitem{ref:edpapero}
E. L. Berger and H. Contopanagos, Phys. Lett. B {\bf 361}, 115 (1995); 
Phys. Rev. {\bf D54}, 3085 (1996); 
ANL-HEP-CP-96-51, hep-ph/9606421 (21 June 1996).
\bibitem{ref:catani}
S. Catani, M. Mangano, P. Nason, and L. Trentadue, Phys. Lett. B {\bf 378}, 329 
(1996); hep-ph/9604351 (18 Apr 1996); M. Mangano and P. Nason, private 
communication.
\bibitem{ref:stermano}
H. Contopanagos and G. Sterman, Nucl. Phys. {\bf B400}, 211 (1993); 
{\bf B419}, 77 (1994).
\bibitem{ref:cteq}
H.L. Lai {\it et al.}, Phys. Rev. {\bf D51} 4763 (1995). 
\bibitem{ref:mrsa}
A. Martin, R. Roberts and W.~J. Stirling, Phys. Lett. B {\bf 354} 155 (1995).
\bibitem{ref:cdfdz} P. Tipton, 28th International Conference on High Energy 
Physics, Warsaw, July, 1996; A. Castro (CDF collaboration), {\it ibid}; 
J. Bantly (D0 collaboration), {\it ibid}.
\bibitem{ref:cdfjets} F. Abe {\it et al}, Phys. Rev. Lett. {\bf 77}, 448 {1996}.

\end{thebibliography}
\end{document}